\documentclass{arxiv}
\usepackage{amssymb}
\usepackage{epsfig}
\usepackage{subfig}
\usepackage{cite}

\begin{document}

\catchline{}{}{}{}{}

\title{The law of the leading digits and the world religions
}

\author{T. A. Mir$^{*}$}

\address{Nuclear Research Laboratory, Astrophysical Sciences Division,\\ Bhabha Atomic Research Centre,\\  Srinagar-190 006, Jammu and Kashmir, India\\
$^{*}$taarik.mir@gmail.com}

\maketitle


\section{Abstract}

Benford's law states that the occurrence of significant digits in many data sets is not uniform but tends to follow a logarithmic distribution such that the smaller digits appear as first significant digits more frequently than the larger ones. We investigate here numerical data on the country-wise adherent distribution of seven major world religions i.e. Christianity, Islam, Buddhism, Hinduism, Sikhism, Judaism and Baha'ism to see if the proportion of the leading digits occurring in the distribution conforms to Benford's law. We find that the adherent data of all the religions, except Christianity, excellently does conform to Benford's law. Furthermore, unlike the adherent data on Christianity, the significant digit distribution of the three major Christian denominations i.e. Catholicism, Protestantism and Orthodoxy obeys the law. Thus in spite of their complexity general laws can be established for the evolution of the religious groups.
\section{Keywords}

Benford's law; religion; adherents 


\section{Introduction}	

In our daily life, we come across huge sets of numerical data being generated due to different activities both human as well as natural. What could be common between two seemingly unrelated data sets like the size of files stored on a computer and the geographical area of the countries? An intuitive reply would be nothing. However, the occurrence of significant digits in many data sets is governed by a common logarithmic distribution which is defined by the equation called Benford's law\cite{Benford}.
\newline
The first significant digit of a number is the first non-zero digit on its extreme left like 5 for 521 and 1 for 0.0184 respectively. The establishment of Benford's law was based on the curious observation made by Simon Newcomb who observed and said that the initial pages of the logarithmic tables were more worn out than the later ones due to the fact that people look more often for numbers with smaller first non-zero digits\cite{Newcomb}. The observation was popularized by Frank Benford (hence the name Benford's law) who then analyzed a large number of unrelated data sets collected from diverse fields e.g. the physical constants, the atomic and molecular masses, the areas of countries, the length of rivers etc., and concluded that the occurrence of first significant digits in all his data sets does follow a logarithmic distribution and hence established the law in the form of following mathematical equation\cite{Benford}. \\ 
\begin{equation}
P(d)= log_{10}(1+\frac{1}{d}), d= 1, 2, 3...,9
\end{equation}
where P(d) is the probability of a number having the first non-zero digit d.\\
Benford's law is also known as ``first digit law'' or ``law of the leading digits''. According to equation (1), in a given data set the probability of occurrence of a certain digit as first significant digit decreases logarithmically as the value of the digit increases from 1 to 9. Thus digit 1 should appear as the first significant digit about $30$\% times, digit 2 about $17$\% times and similarly 9 should appear about $4$\% times. This observation is in stark contrast to the equal probability of about $\frac{1}{9}$ =$11$\% for each digit from 1 to 9 if all are equally likely to occur as the first significant digit. 
\newline
Despite many of its properties like base and scale invariance being unraveled\cite{Hill, Hill1} over more than a century after its discovery a complete explanation of Benford's law has remained elusive\cite{Berger}. However, increasing number of data sets from unrelated phenomena have been tested to conform to this law with high statistical accuracy. After remaining obscure for decades, the law is finding wide spread applications. In economics, it has been validated for stock market prices\cite{Pietronero} and applied in detecting the fraudulent data submitted by the companies\cite{Nijrini}, verifying the reliability of macroeconomic data of the countries\cite{Nye} and in the determination of winning bids in e-Bay auctions\cite{Giles}. In physics the law has been successfully applied to numerical data on physical constants\cite{Burke}, atomic spectra\cite{Pain}, decay width of hadrons\cite{Shao}, magnitude and depth of earthquakes\cite{Pietronero, Sambridge} and mantissa distribution of pulsars\cite{Shao1}. The law has been utilized in biological sciences to check the veracity of the data on clinical trials and discovery of drugs\cite{Beer, Orita}, study of diseases and genes\cite{Hernandez}. The law is used in optimizing the size of computer files\cite{Barlow}, enhancing the computing speed\cite{Schatte} and appearance of numbers on the internet\cite{Dorogovtsev}. On the social sciences front the law has been used to detect election frauds and anomalies\cite{Mebane}.
\newline
We show here the applicability of Benford's law to the field of religious demography. Though very important, the research on religious demography is still in infancy and admittedly inconsistencies in world religion adherent data cannot be ruled out completely. Besides reshaping the social landscape, the change in religious demographics is also affecting the political decisions of various countries. To prevent problems like making a flawed policy decision due to the use of some erroneous data it is the obligation of the researchers to reassure the policy makers about the quality of the data. We analyze the available data on the country-wise adherent distribution of seven major world religions Christianity, Islam, Buddhism, Hinduism, Sikhism, Judaism and Baha'ism. We find that country-wise adherent data on all religions except Christianity follow Benford's law. We also reveal the validity of Benford's law on Catholicism, Protestantism and Orthodoxy, the three major denominations of Christianity.

\section{Data}
Due to the developmental stage of the religious demography research, obtaining complete data on  adherents of a particular faith and its distribution from a single source is not quite easy. The Wikipedia combines the available data from different sources like the CIA World Factbook, US State Department's International Religious Freedom Report, Pew Research Center, Adherents.com, Joshua Project etc. to arrive at comprehensive data tables for the country-wise adherent distribution of major world religions. The data on country-wise number of Muslims are taken from Pew Research Center study on the size and distribution of world's Muslim population\cite{Pew, Grim1}. The number of adherents of Baha'i faith in each country is based on the 2005 estimates of World Christian Database as reported in The Association of Religion Data Archive (ARDA)\cite{Baha'i}.

\subsection{Data analysis and Results}

The detailed statistical analysis of the country-wise adherent distribution is shown in Tables 1 and 2. The $N_{Obs}$, the number of counts appearing in the corresponding data set as first significant digit, against each digit from 1 to 9 are shown for each religion in columns 2, 3, 4, 5, 6 of Tables 1 and 2. We also show $N_{Ben}$, the corresponding counts (in brackets) for each digit as predicted by Benford's law: 
\begin{equation}
N_{Ben}= N log_{10}(1+\frac{1}{d})
\end{equation} 
along with the root mean square error ($\Delta{N}$) calculated from the binomial distribution\cite{Shao}
\begin{equation}
\Delta{N}= \sqrt{NP(d)(1-P(d))}
\end{equation} 
where $N$ for each religion is the total number of countries its adherents are reported to exist. For example, as shown in column 2 of Table 1, the adherents of Christianity are reported in 205 territories\cite{Wiki}. The observed count for digit 1 as first significant digit is 50 whereas the expected count from Benford's law is 61.7 with an error of about 6.6. To evaluate the degree of agreement between the observed and expected first digit distributions let us first write the \textit{Null Hypothesis}, $H_{O}$ that the observed distribution of the first significant digit in each of the case we consider is same as expected on the basis of Benford's law. To test the null hypothesis we carry out the Pearson's $\chi^{2}$ test. 
\begin{equation}
  \chi^{2}(n-1) =\sum_{i=1}^n\dfrac{(N_{Obs}-N_{Ben})^{2}}{N_{Ben}}
\end{equation}

In our case $n=9$ which means we have $n-1=8$ degrees of freedom. Under $95\%$ confidence level (CL) $\chi^{2}(8)$=15.507 which is the critical value for acceptance or rejection of null hypothesis. If the value of the calculated $\chi^{2}$ is less than the critical value then we accept the null hypothesis and conclude that the data fits Benford's law.
\newline
Christianity has the largest number of adherents spread across 205 countries (column 2 of Table 1). The $\chi^{2}$= 16.419 (the last row and column 2 of Table 1) is larger than $\chi^{2}(8)=15.507$ and thus forcing us to reject the null hypothesis indicating in turn that the data on the country-wise adherent distribution of Christianity does not follow Benford's law. We further checked the behavior of three largest Christian denominations, Catholicism (197 countries)\cite{Wiki1}, Protestantism (171 countries)\cite{Wiki2} and Orthodoxy (42 countries)\cite{Wiki3} as shown in Columns 3, 4 and 5. Surprisingly their respective $\chi^{2}$ values 10.143, 5.208, 6.946 turn out to be lower than $\chi^{2}(8)=15.507$ and hence null hypothesis must be accepted. In column 6 of Table 1 we show the counts of significant digits of Muslim populations from 184 countries as reported in a comprehensive demographic study of Muslims conducted by Pew Research Center\cite{Pew, Grim1}. The Pearson's $\chi^{2}=10.646$ is well below the critical value of $15.507$ and null hypothesis is accepted. In Table 2 we show the data on Baha'ism (175 countries)\cite{Baha'i}, Buddhism (129 countries)\cite{Buddhism}, Hinduism (97 countries)\cite{Hinduism}, Sikhism (47 countries)\cite{Sikhism} and Judaism (107 countries)\cite{Judaism}. The smaller values of the Pearson's $\chi^{2}$ make the null hypothesis acceptable and hence Benford's law describes the distribution of significant digits in each case. The Wikipedia table on the number of adherents of Buddhism shows two values for some countries. Accordingly we consider both the situations taking into account the lower and the upper values separately. When smaller values are counted for the Buddhism (not shown here), the Pearson's $\chi^{2}$ $6.633$ turns out to be smaller than the critical value.  The lowest value of $\chi^{2}$ $1.786$ occurs for Buddhism when only the larger values are considered for countries where two values are given. Further, although the sample size of the adherent data on Sikhism is quite small, reported only in 47 countries, the $\chi^{2}=6.863$ indicates that Benford's law upholds. Finally in Table 3 we sum up the number of counts for significant digits appearing in the adherent data sets on all the religions considered in our study. The Pearson's $\chi^{2}$=7.310 being smaller than the critical value 15.507 shows that Benford's law describes the adherent data of major world religions with a reasonable accuracy.
The results reported in the tables can readily be appreciated from Figs. 1-3. It can be seen that the occurrence of the significant digits in country-wise adherent distribution of religions closely resembles those predicted by Benford's law.
\newpage
\begin{table}[ph]
\tbl{ The significant digit distribution of country-wise Christian, Catholic, Protestant, Orthodox and Muslim populations}
{\begin{tabular}{@{}llllll@{}} \toprule
First Digit & Christian  & Catholic & Protestant & Orthodox & Muslim \\
& (205) & (197) & (171) & (42) & (184)\\ \colrule
$1$ \hphantom{00} & 50 (61.7$\pm$6.6) & 47 (59.3$\pm$6.4)& 48 (51.5$\pm$6.0)& 12 (12.6$\pm$3.0)&60 (55.4$\pm$6.2)\\
$2$ \hphantom{00} & 30 (36.1$\pm$5.4) & 34 (34.7$\pm$5.3)& 25 (30.1$\pm$5.0)& 5 (7.3$\pm$2.5)& 43 (32.4$\pm$5.2)\\
$3$ \hphantom{00} & 29 (25.6$\pm$4.7) & 31 (24.6$\pm$4.6)& 26 (21.4$\pm$4.3)& 7 (5.2$\pm$2.1)& 21 (23.0$\pm$4.5) \\
$4$ \hphantom{00} & 27 (19.9$\pm$4.2) & 18 (19.1$\pm$4.1)& 15 (16.6$\pm$3.9)& 3 (4.1$\pm$2.0) & 13 (17.8$\pm$4.0) \\ 
$5$ \hphantom{00} & 14 (16.2$\pm$3.9) & 19 (15.6$\pm$3.8)& 18 (13.5$\pm$3.5)& 4 (3.3$\pm$1.7) & 7  (14.6$\pm$3.7) \\
$6$ \hphantom{00} & 15 (13.7$\pm$3.6) & 10 (13.2$\pm$3.5)& 10 (11.4$\pm$3.3)& 4 (2.8$\pm$1.6) & 11 (12.3$\pm$3.4) \\
$7$ \hphantom{00} & 18 (11.8$\pm$3.3) & 10 (11.4$\pm$3.3)& 9 (9.9$\pm$3.0)& 5 (2.4$\pm$1.5) & 13 (10.7$\pm$3.2) \\
$8$ \hphantom{00} & 6  (10.5$\pm$3.1) & 15 (10.1$\pm$3.1)& 12 (8.7$\pm$2.9)& 2 (2.1$\pm$1.4) &10  (9.4$\pm$3.0) \\
$9$ \hphantom{00} & 16 (9.4$\pm$3.0)    & 13 (9.0$\pm$2.9)& 8 (7.8$\pm$2.7)& 0  (1.9$\pm$1.3) &6 (8.4$\pm$2.8)\\ 
\botrule
\textbf{Pearson} $\chi^{2}$ \hphantom{00} & \bf16.419 &  \bf10.143 & \bf 5.208 & \bf 6.946  & \bf\bf10.646 \\ \botrule
\end{tabular} \label{ta1}}
\end{table}

\begin{table}[ph]
\tbl{ The significant digit distribution of country-wise Baha'i, Buddhist, Hindu, Sikh and Jewish populations}
{\begin{tabular}{@{}llllll@{}} \toprule
First Digit & Baha'i & Buddhist & Hindu & Sikh & Jew\\
& (175) & (129) & (97) & (47) &  (107)\\ \colrule
$1$ \hphantom{00} & 61 (52.9$\pm$6.1) &  38 (38.8$\pm$5.2)& 26 (29.2$\pm$4.5)& 12 (14.1$\pm$3.1)&  42 (32.2$\pm$4.7)\\
$2$ \hphantom{00} & 35 (30.8$\pm$5.0) & 21 (22.7$\pm$4.3)& 16 (17.1$\pm$3.7)& 11 (8.3$\pm$2.6) & 13 (18.8$\pm$4.0) \\
$3$ \hphantom{00} & 20 (21.9$\pm$4.4) & 15 (16.1$\pm$3.7)& 12 (12.1$\pm$3.2)& 6 (5.9$\pm$2.3) & 15 (13.4$\pm$3.4) \\
$4$ \hphantom{00} & 11 (16.9$\pm$3.9) & 15 (12.5$\pm$3.4)& 10 (9.4$\pm$2.9)& 2 (4.5$\pm$2.0) & 6 (10.4$\pm$3.1)\\ 
$5$ \hphantom{00} & 10  (13.8$\pm$3.6) & 10 (10.2$\pm$3.1)& 8 (7.7$\pm$2.6)& 5 (3.7$\pm$1.8) & 12 (8.5$\pm$2.8)\\
$6$ \hphantom{00} & 16 (11.7$\pm$3.3) & 8 (8.6$\pm$2.8)& 11 (6.5$\pm$2.5)& 1 (3.1$\pm$1.7) & 7 (7.2$\pm$2.6)\\
$7$ \hphantom{00} & 9  (10.1$\pm$3.1) & 10 (7.5$\pm$2.6)& 3 (5.6$\pm$2.3)& 5 (2.7$\pm$1.6) & 6 (6.2$\pm$2.4)\\
$8$ \hphantom{00} & 5  (8.9$\pm$2.9) & 7 (6.6$\pm$2.5)& 3 (5.0$\pm$2.1)& 2 (2.4$\pm$1.5) & 3 (5.5$\pm$2.3)\\
$9$ \hphantom{00} & 8 (8.0$\pm$2.8)  & 5 (6.0$\pm$2.4)& 8 (4.4$\pm$2.0)& 3  (2.1$\pm$1.4) & 3 (4.9$\pm$2.2)\\ 
\botrule
\textbf{Pearson} $\chi^{2}$ \hphantom{00} & \bf8.648 &  \bf1.786 & \bf8.45 & \bf 6.863  & \bf10.158 \\ \botrule
\end{tabular} \label{ta1}}
\end{table}

\begin{table}[ph]
\tbl{ The significant digit distribution of country-wise adherents of all religions}
{\begin{tabular}{@{}lllllllllll@{}} \toprule
First Digit & 1 & 2 & 3 & 4 & 5 & 6 & 7 & 8 & 9 & Total\\
\\ \colrule
$N_{Obs}$ \hphantom{00} & 289  & 169 & 118 & 84 &  66  & 69 & 64 & 36 & 49 & \bf944\\
$N_{Ben}$ \hphantom{00} & 284.2  & 166.2 & 117.9 & 91.5 & 74.7 & 63.2 & 54.7 & 48.3 & 43.2 \\
Error \hphantom{00} & 14.1 & 11.7 & 10.2 & 9.1 & 8.3 & 7.7  & 7.2 & 6.8 & 6.4  \\
\botrule
\textbf{Pearson} $\chi^{2}$ \hphantom{00} & &&&&&&&&&\bf7.766\\ \botrule
\end{tabular} \label{ta1}}
\end{table}

\begin{figure}
\begin{minipage}[b]{.9\linewidth}
\vspace*{-5pt}
\hspace*{5pt}
\centering
\begin{tabular}{cc}
\hspace*{10pt}
\vspace*{-70pt}
\epsfig{file=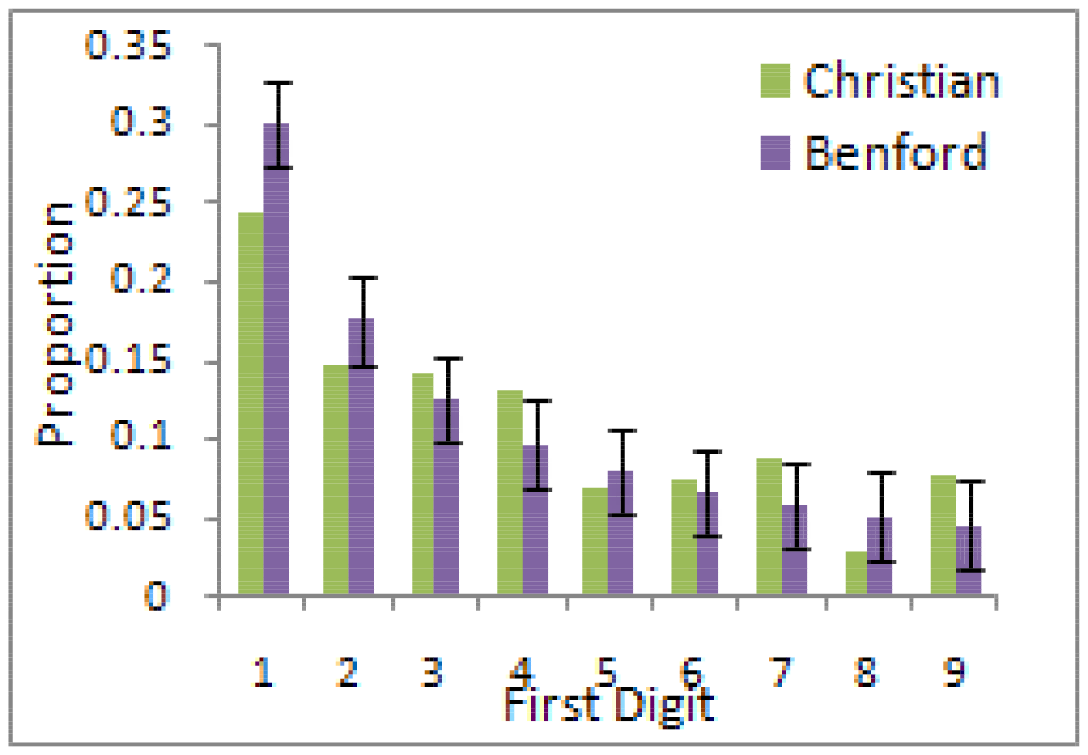,width=0.7\linewidth, height=0.9\linewidth, angle=270, clip=}&
\hspace*{-140pt}
\epsfig{file=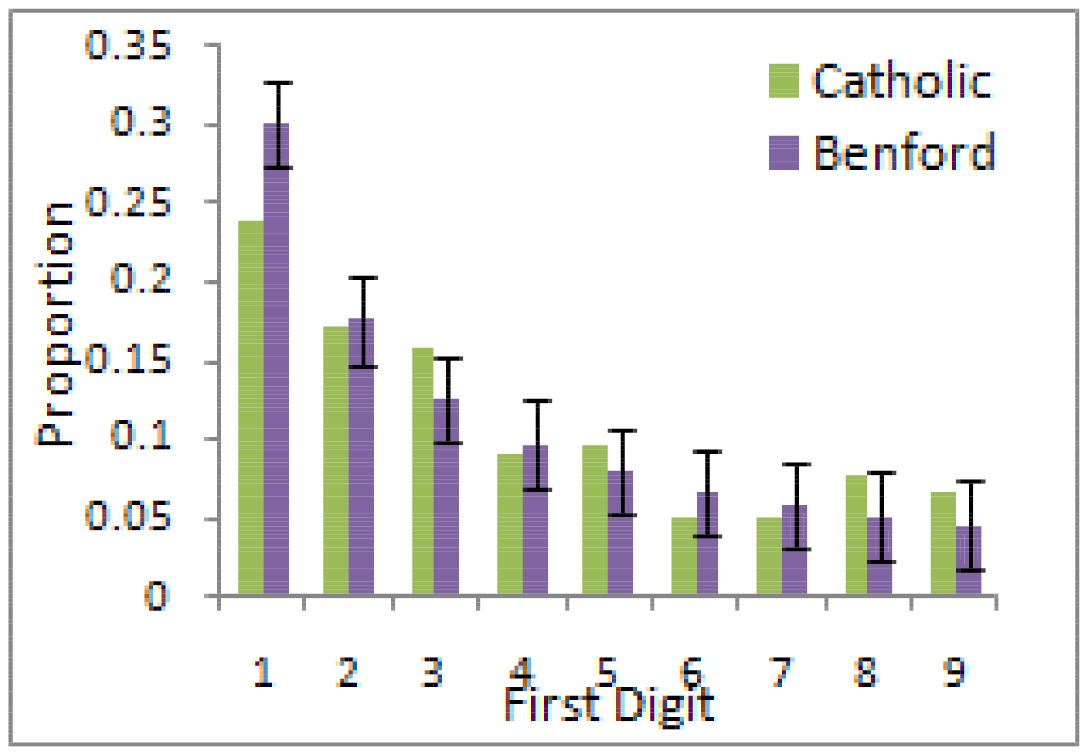,width=0.7\linewidth, height=0.9\linewidth, angle=270,  clip=} \\
\hspace*{10pt}
\vspace*{-70pt}
\epsfig{file=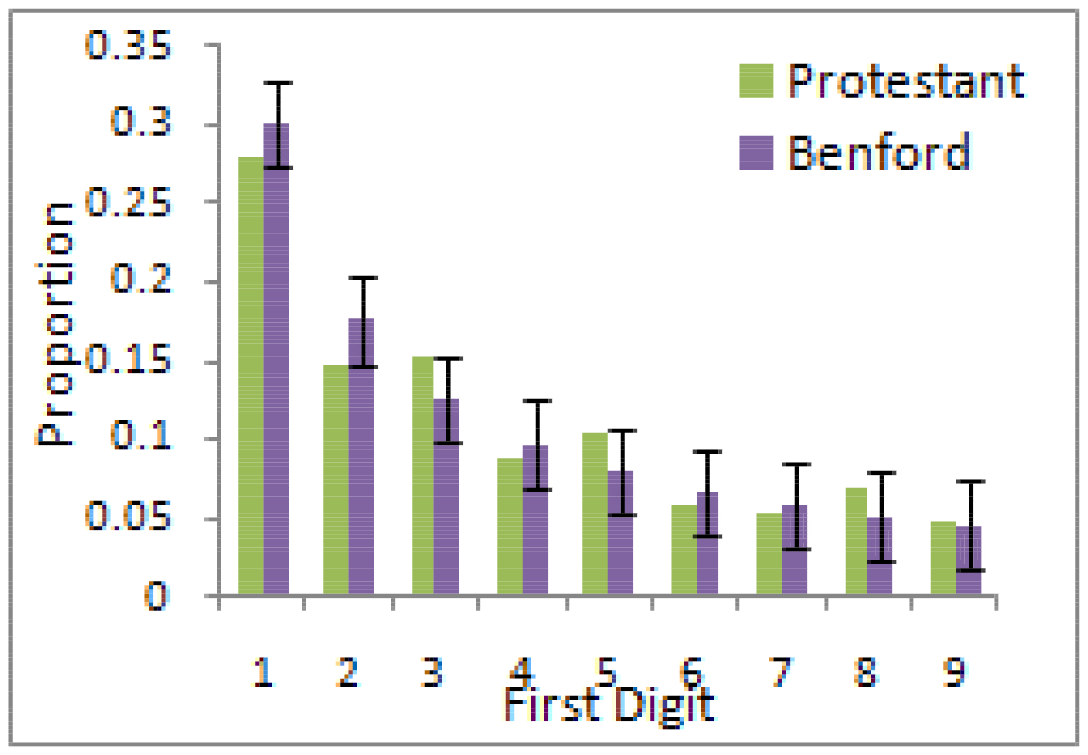,width=0.7\linewidth,height=0.9\linewidth, angle=270, clip=} &
\hspace*{-140pt}
\epsfig{file=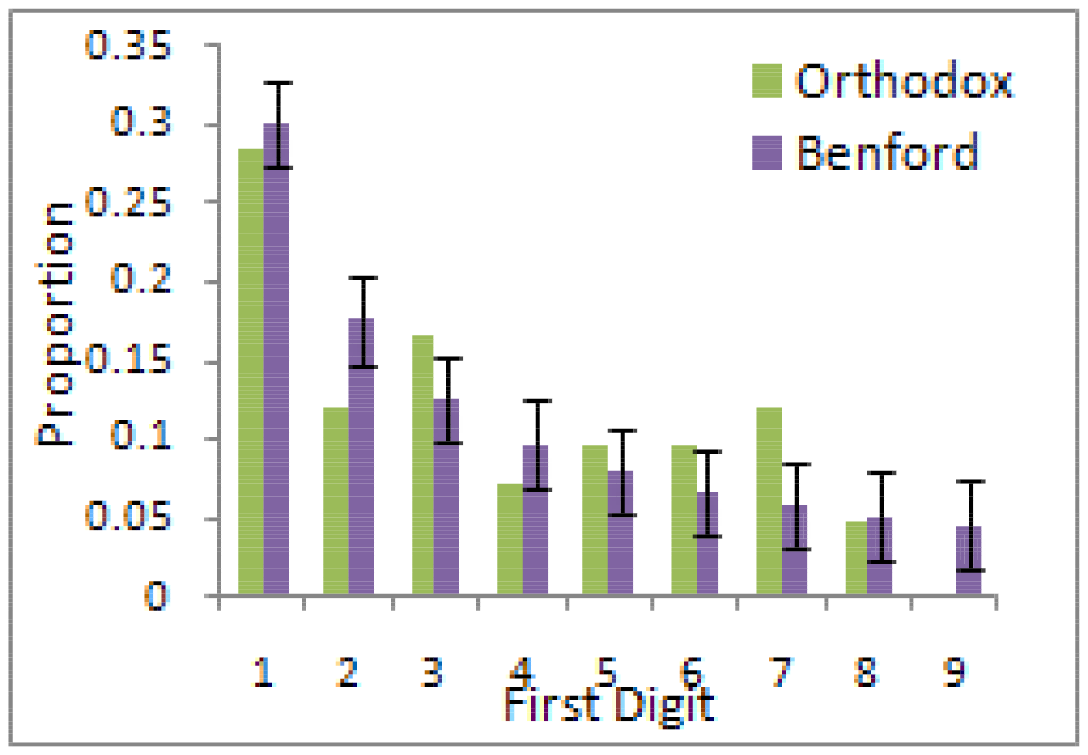,width=0.7\linewidth, height=0.9\linewidth, angle=270, clip=}\\
\hspace*{10pt}
\vspace*{-70pt}
\epsfig{file=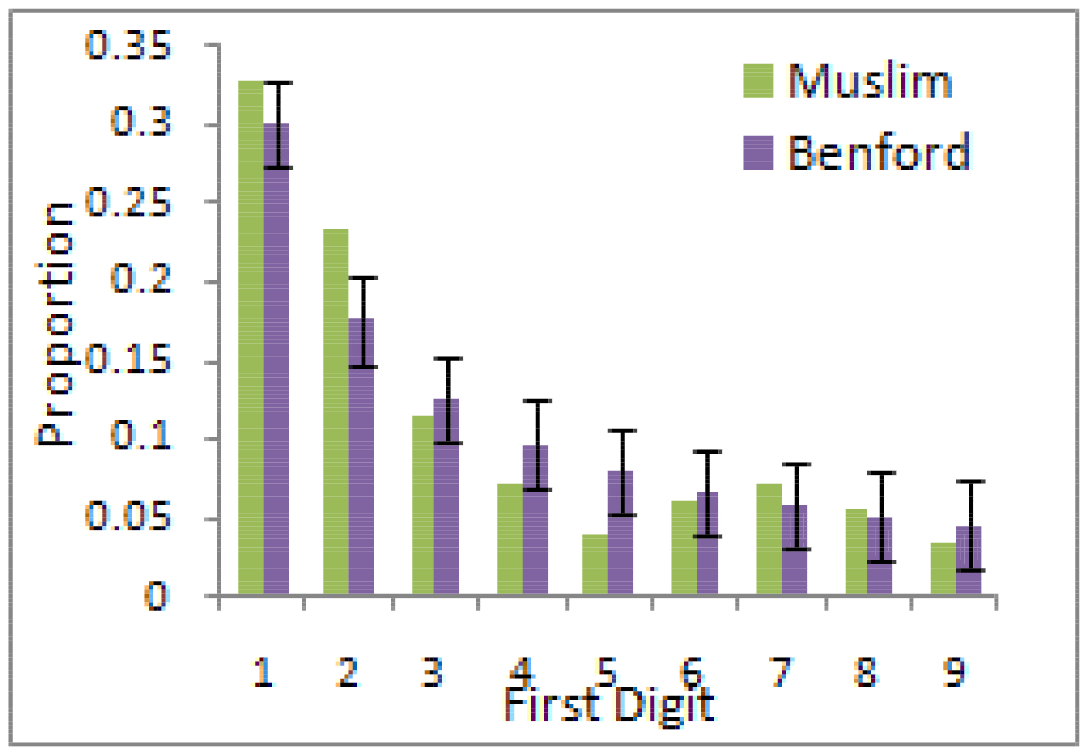,width=0.7\linewidth, height=0.9\linewidth, angle=270, clip=} &
\hspace*{-140pt}
\epsfig{file=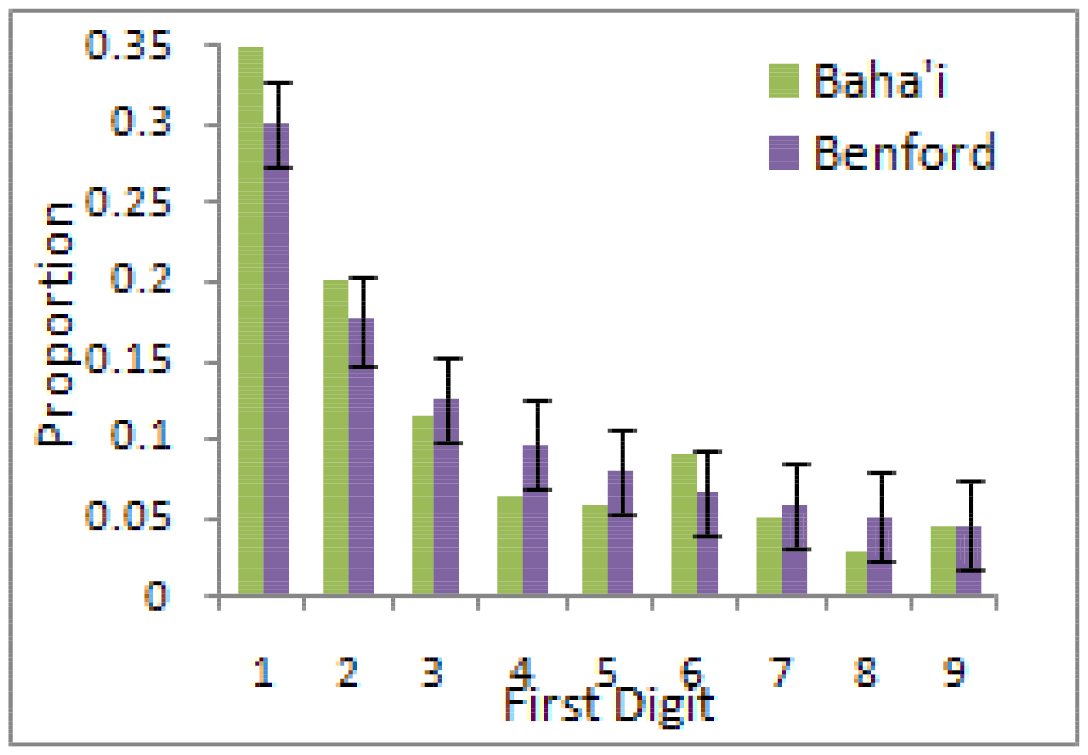,width=0.7\linewidth, height=0.9\linewidth, angle=270, clip=}\\
\end{tabular}
\vspace*{10pt}
\hspace*{20pt}
\end{minipage}
\caption{Observed and Benford distributions of significant digits for country-wise adherents of Christianity, its three major denominations i.e. Catholicism, Protestantism and Orthodoxy, Islam and Baha'ism}
\end{figure}

\begin{figure}
\begin{minipage}[b]{.9\linewidth}
\vspace*{-5pt}
\hspace*{-10pt}
\centering
\begin{tabular}{cc}
\hspace*{10pt}
\vspace*{-70pt}
\epsfig{file=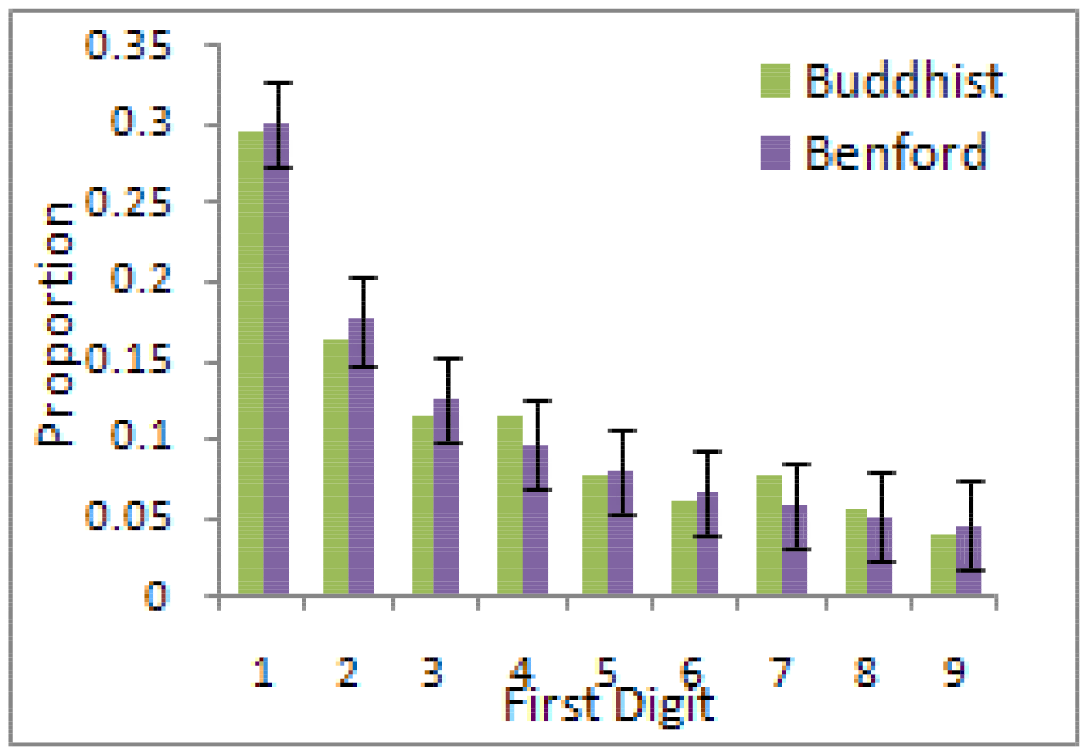,width=0.7\linewidth, height=0.9\linewidth, angle=270,clip=} &
\hspace*{-140pt}
\epsfig{file=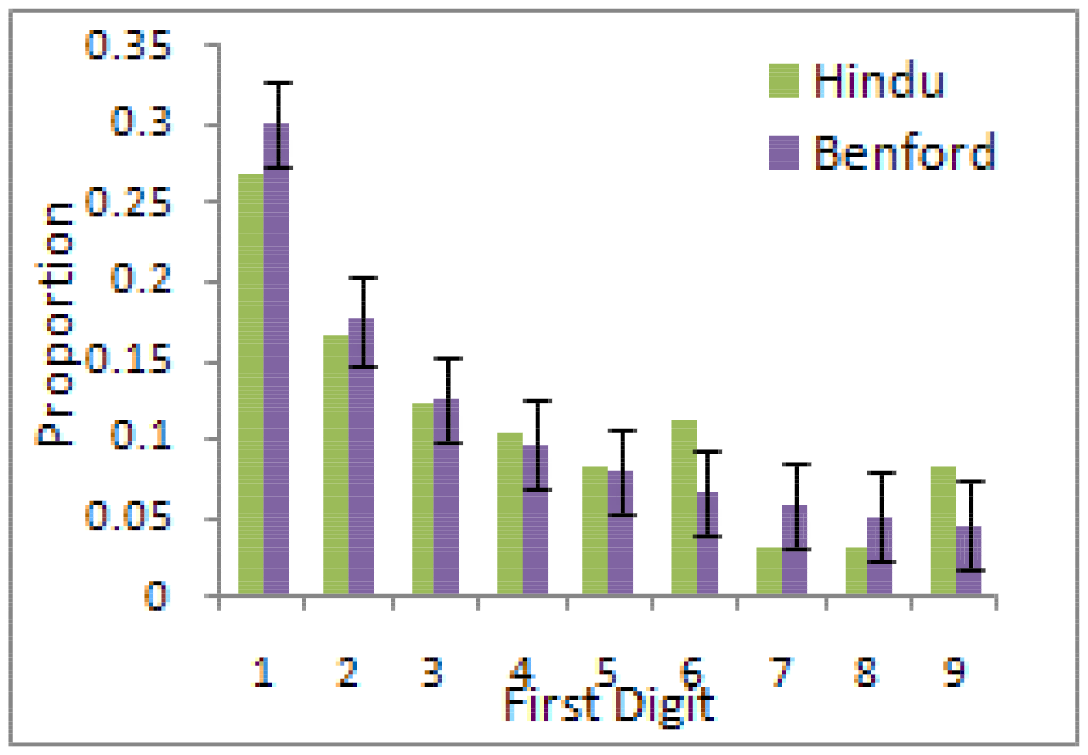,width=0.7\linewidth, height=0.9\linewidth, angle=270, clip=}\\
\hspace*{10pt}
\vspace*{-70pt}
\epsfig{file=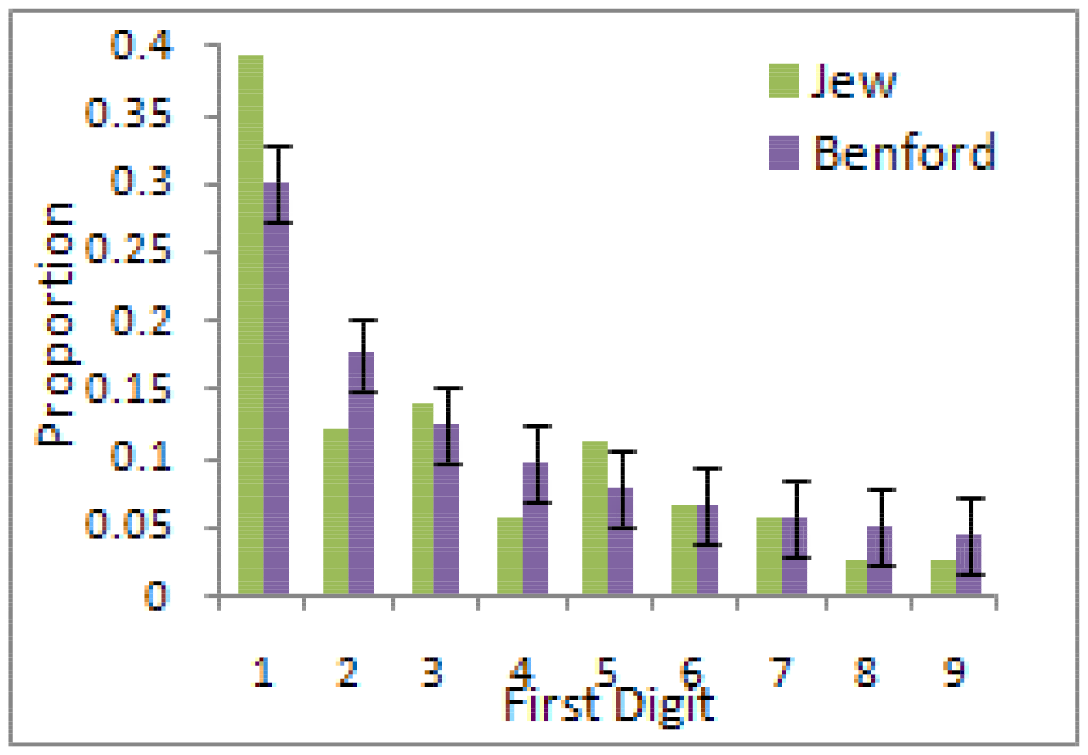,width=0.7\linewidth, height=0.9\linewidth, angle=270, clip=} &
\hspace*{-140pt}
\epsfig{file=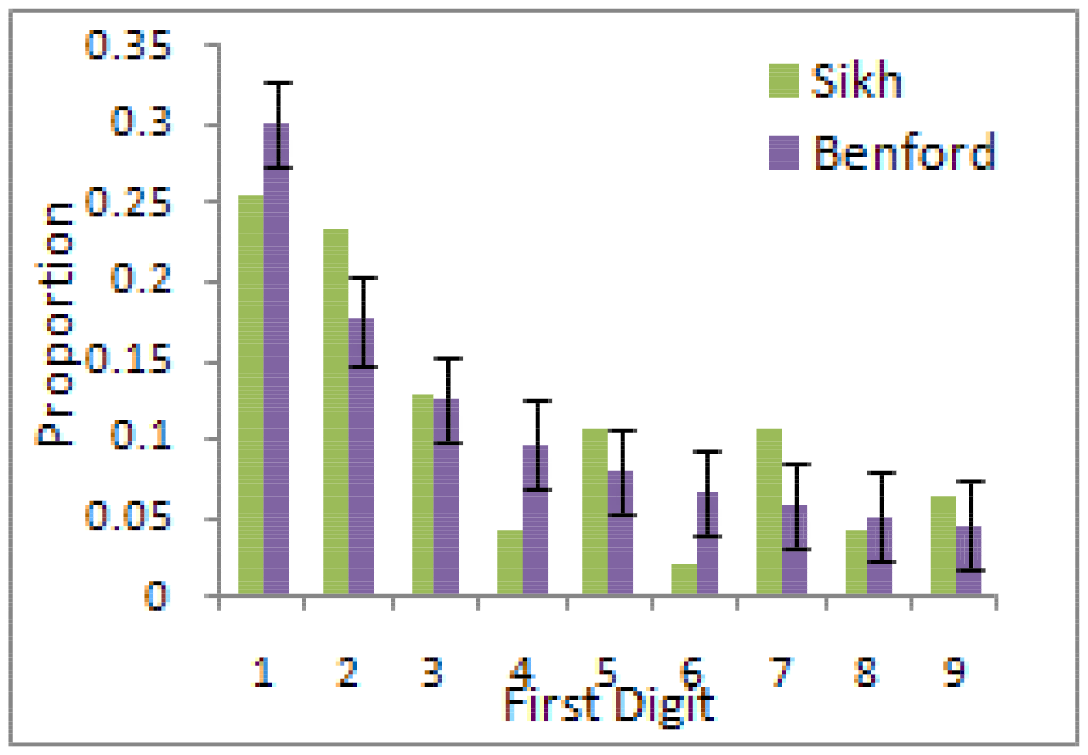,width=0.7\linewidth, height=0.9\linewidth, angle=270, clip=}\\
\end{tabular}
\vspace*{10pt}
\hspace*{20pt}
\end{minipage}
\caption{Observed and Benford distributions of significant digits for country-wise adherents of Buddhism, Hinduism, Judaism and Sikhism}
\end{figure}


\begin{figure}

\centering
\hspace*{50pt}
\epsfig{file=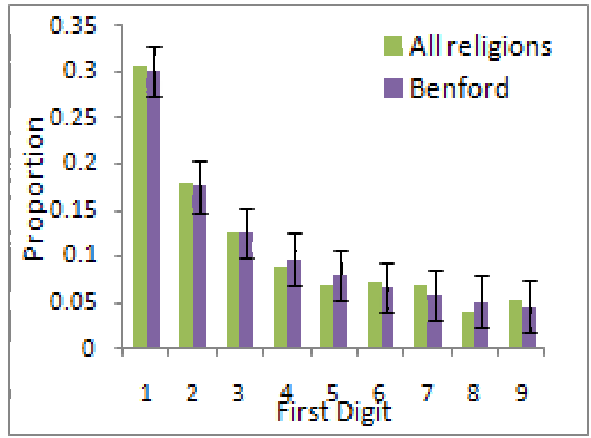,width=0.7\linewidth, height=0.9\linewidth, angle=270, clip=} 
\vspace*{-60pt}
\caption{Observed and Benford distributions of significant digits for country-wise adherents of all religions}
\end{figure}

In the case of any major deviations from the predictions of Benford's law, the distributions as shown in figures could hardly be revealed. Particularly from Fig. 3 it is clear that agreement of the observed proportion of occurrence of significant digits for all the religions to those predicted by Benford's law is excellent.
\section{Discussion}

Despite being a key social attribute and the largest scale phenomenon involving people beyond the geographical boundaries of countries, the adherence to a religion, has received relatively lesser scientific attention\cite{Johnson}. Furthermore, religion being a dominant force in society for both cohesion and conflict\cite{Grim}, when it comes to the scientific study of religions the most debated question, having serious social and political consequences, is the estimation of size distribution of different religious groups\cite{Grim1, Foreign, Hsu}. Studies have shown that the growth dynamics of religious activities are similar to those found in economic and scientific activities\cite{Picoli}. The adherent distribution of major world religions has been shown to obey Zipf's and Pareto distributions and their evolution has been described using the same equations as for the crystallization process\cite{Ausloos, Ausloos1}. A recent mathematical model has predicted the extinction of religion in societies which perceive the utility of not adhering to be greater than the utility of adhering\cite{Abrams}. 
\newline  
Here we have shown that populations of major world religions follow Benford's law. It is worthwhile to mention here that F. Benford used population size of U.S. towns as input data to test and examine the accuracy of his observation\cite{Benford}. Later on, the population of the 198 countries of world have been shown to follow the first digit law\cite{Sandron}. However, being undertaken by the individual governments, the census estimates of the total populations of countries are expected to be reliable. Same is not true for the estimates of adherents of different religions. Several religions compete with each other to increase the number of their adherents. Such followers can either be converted from one religion to another or become followers of a particular religion though previously being associated with none\cite{Vitanov}. This continuous evolution of religions puts serious restraints on determining the exact number of adherents of a particular religion, a task which is further complicated due to government favor to state religions in certain countries, government restriction on religious census or biased data from different religious associations for political propaganda\cite{Usreport}. Thus errors are expected to be found in the data on religious demographics and data must be subjected to scientiﬁc scrutiny for any alterations.   
\newline
Benford's law holds for those data sets in which the occurrence of numbers is free from any restrictions. It has been found that tampered, unrelated or fabricated numbers usually do not follow Benford's law\cite{Hill2}. Significant deviations from the Benford distribution may indicate fraudulent or corrupted data\cite{Nijrini1}. In the light of these issues it would be interesting to check if the adherent data of major world religions submit to Benford's law and any departure from the law may alert the researcher about possible data misrepresentation.
\newline 
In our study we observed that the adherent data of three major Christian denominations follow Benford's law. However, when Christianity is considered as a single religious group, the distribution of the significant digits of the adherent data deviates from the predictions of Benford's law. The deviation could possibly arise from the incorrect method used or erroneous grouping of data from different sources in Wikipedia. Further, the deviation from the Benford distribution may also arise from overestimation/underestimation of the number of Christian adherents in some countries. The absence of a single religion database has forced many people including scholars to turn to Wikipedia for information on major world religions. But relevant Wikipedia data has been assumed to be devoid of sufficient academic rigor and consistency\cite{Grim1}. However, the prevalence of Benford's law in the adherent data on three largest Christian denominations, Buddhism, Hinduism, Sikhism, Judaism and Baha'ism infuses some confidence in the quality of the data available on Wikipedia and the method employed to collect and combine the data from different sources to arrive at these comprehensive adherent data tables.
Now in the last decade, Muslim societies have generated considerable public interest and their population has come under increased scrutiny\cite{Pew, Grim1}. There are varying estimates of the size of world's Muslim population. We analyzed the estimates of the global Muslim population reported in the most recent study done by the Pew Research Center and found a reasonable agreement to Benford's law which in turn indicates the reliability of this demographic survey. The validity of Benford's law for adherent distribution of major world religions is a further hint that universal growth mechanisms might exist which can give rise to general laws independent of the particular details of the systems\cite{Picoli}. 
\section{Conclusion}
The religion is the largest scale phenomenon in a society that has a bearing on social, economical and political behavior of its adherents. We applied, for the first time, Benford's law to the data on the adherent distribution of seven major world religions. We found that the adherent data of Islam, Hinduism, Buddhism, Sikhism, Judaism and Baha'ism follow Benford's law. Similarly, the adherent data of the three major Christian denominations Catholicism, Protestantism and Orthodoxy also follow the law but contrarily cumulative adherent data of Christianity when considered as a single religious group does not. Notwithstanding the complex path taken by a religious group to attain a certain state, its macroscopic behavior satisfies the simple law of leading digits.

\section*{Acknowledgments}
Suggestions from M. Ausloos and P. M. Ishtiaq are gratefully acknowledged.  



\end{document}